\documentclass[twocolumn,notitlepage,nofootinbib]{revtex4-2}
\usepackage{amsmath,amssymb,bm,graphicx,hyperref}
\allowdisplaybreaks[1]

\newcommand\<{\langle}
\renewcommand\>{\rangle}
\newcommand\eps{\epsilon}
\renewcommand\r{\bm{r}}
\newcommand\R{\bm{R}}
\newcommand\Z{\mathbb{Z}}
\newcommand\sol{\mathrm{sol}}
\DeclareMathOperator\arccot{arccot}
\let\Re\relax\DeclareMathOperator\Re{Re}
\let\Im\relax\DeclareMathOperator\Im{Im}

\begin{document}

\title{Universal bound states and resonances with Coulomb plus short-range potentials}

\author{Shunta Mochizuki}
\author{Yusuke Nishida}
\affiliation{Department of Physics, Tokyo Institute of Technology,
Ookayama, Meguro, Tokyo 152-8551, Japan}

\date{August 2024}

\begin{abstract}
We study charged particles in three dimensions interacting via a short-range potential in addition to the Coulomb potential.
When the Bohr radius and the scattering length are much larger than the potential range, low-energy physics of the system becomes independent from details of the short-range potential.
We develop the zero-range theory to describe such universal physics in terms of the Bohr radius and the scattering length by generalizing the Bethe-Peierls boundary condition, which is then applied to two charged particles to reveal their bound states and resonances.
Infinite resonances are found for a repulsive Coulomb potential, one of which turns into a bound state with increasing inverse scattering length, whereas infinite bound states exist for an attractive Coulomb potential with no resonances at any scattering length.
The zero-range theory is also applied to three equally charged particles at infinite scattering length under the variational Born-Oppenheimer approximation.
We find that the effective potential between two heavy particles induced by a light particle is an inverse-square attraction at distances shorter than the Bohr radius, leading to infinite deep bound states, whereas shallow ones successively turn into resonances with increasing Coulomb repulsion.
\end{abstract}

\maketitle

\section{Introduction}
When particles interact via a short-range potential with its scattering length much larger than the potential range, low-energy physics of the system becomes universal, i.e., independent from details of the short-range potential~\cite{Braaten:2006}.
Such universal physics is parametrized by the scattering length only and can be relevant to diverse systems in physics due to the universality.
One representative example is the BCS-BEC crossover for many fermions~\cite{Bloch:2008,Giorgini:2008,Zwerger:2012,Randeria:2014}, which has attracted considerable interest not only from ultracold atoms but also from nuclear physics~\cite{Gandolfi:2015,Horikoshi:2019}.
Another representative example is the Efimov effect for few bosons~\cite{Nielsen:2001,Jensen:2004,Braaten:2006,Braaten:2007,Hammer:2010,Ferlaino:2011,Naidon:2017,Greene:2017,D'Incao:2018}, which was theoretically predicted in nuclear physics~\cite{Efimov:1970,Efimov:1973}, experimentally observed in ultracold atoms~\cite{Kraemer:2006}, and potentially expected in condensed-matter systems~\cite{Nishida:2013,Kato:2020,Nakayama:2021}.

A theoretical framework to describe the universal physics in terms of the scattering length is called the zero-range theory, which can be implemented in various ways~\cite{Braaten:2006}.
One naive way is to take the zero-range limit of some finite-range potential under fixed scattering length, whereas a zero-range potential can directly be constructed with the Huang-Yang pseudopotential being the most famous among other equivalent representations~\cite{Huang:1957,Olshanii:2001}.
Instead of providing the Hamiltonian with potentials, the zero-range theory can also be implemented by imposing the so-called Bethe-Peierls boundary condition on the wave function,
\begin{align}\label{eq:bethe-peierls}
\lim_{r\to0}\psi(r) \propto \frac1r - \frac1a + O(r),
\end{align}
allowing the singularity determined by the scattering length $a$~\cite{Bethe:1935}.

The above zero-range theory assumes a short-range potential only, so that it cannot be applied to charged particles interacting via the Coulomb potential in addition to a short-range potential.
Such charged particles and resulting bound states and resonances, however, have been important in diverse fields such as atomic, molecular, and chemical physics~\cite{Domcke:1981,Domcke:1983,Florescu-Mitchell:2006,Lucchese:2019}, nuclear and hadron physics~\cite{Bethe:1949,Jackson:1950,Blatt-Weisskopf:1952,Taylor:1972,Kong:1999,Kong:2000,Higa:2008,Konig:2013,Hammer:2017}, and dark-matter astrophysics~\cite{Hisano:2003,Hisano:2004,Hisano:2005,Arkani-Hamed:2009,Braaten:2017,Braaten:2018a,Braaten:2018b}.
Therefore, it is worthwhile to develop the zero-range theory in the presence of a Coulomb potential and reveal universal aspects of charged particles interacting via a short-range potential.
To this end, we first present the zero-range theory in Sec.~\ref{sec:zero-range} of this paper, which is then applied to two charged particles in Sec.~\ref{sec:two-particles} and three charged particles in Sec.~\ref{sec:three-particles}.
Several examples from nuclear and hadronic systems are also discussed in Sec.~\ref{sec:repulsive}, and our work is finally summarized in Sec.~\ref{sec:summary} together with some outlook.

\section{Zero-range theory}\label{sec:zero-range}
\subsection{Self-adjoint extension}
Let us study two charged particles in three dimensions, whose relative wave function in the $s$-wave channel obeys $\hat{H}r\psi(r)=Er\psi(r)$
with
\begin{align}\label{eq:radial}
\hat{H} = -\frac{\hbar^2}{2m}\frac{d^2}{dr^2} \pm \frac{\hbar^2}{ma_0r}.
\end{align}
Here, $r\in(0,\infty)$ is the interparticle separation, $m$ is the reduced mass, $a_0$ is the Bohr radius, and the upper (lower) sign corresponds to the repulsive (attractive) Coulomb potential throughout this section.
We wish to implement the zero-range interaction by finding appropriate boundary conditions on the wave function at $r\to0$, which can be achieved based on the self-adjoint extension of the Hamiltonian.
Because its detailed accounts can be found, for example, in Refs.~\cite{Bonneau:2001,Gitman-Tyutin-Voronov:2012,Teschl:2014}, only the essentials are described below.

We introduce arbitrary $\chi(r)=\sqrt{4\pi}r\psi(r)$ and $\varphi(r)$ assumed to be square integrable on $(0,\infty)$.
The Hamiltonian has to be Hermitian, meaning that the right-hand side of
\begin{align}\label{eq:hermitian}
& \int_0^\infty\!dr\,\varphi(r)^*\hat{H}\chi(r)
- \int_0^\infty\!dr\,[\hat{H}\varphi(r)]^*\chi(r) \notag\\
&= \frac{\hbar^2}{2m}\lim_{r\to0}W[\varphi(r)^*,\chi(r)]
\end{align}
vanishes with $W[\varphi(r)^*,\chi(r)]=\varphi(r)^*\chi'(r)-\varphi'(r)^*\chi(r)$ being the Wronskian.
Furthermore, the Hamiltonian has to be self-adjoint, meaning that the maximal domain of $\varphi$ on which $\hat{H}$ can act coincides with the given domain of $\chi$.
For example, if $\lim_{r\to0}\chi(r)=0$ and $\lim_{r\to0}\chi'(r)=0$ are imposed, $\hat{H}$ is Hermitian without any conditions on $\varphi$, violating its self-adjointness.
On the other hand, if only $\lim_{r\to0}\chi(r)=0$ is imposed, $\lim_{r\to0}\varphi(r)=0$ is required in order for $\hat{H}$ to be Hermitian, making it self-adjoint as well.
The latter case is the usual boundary condition with no zero-range interaction implemented.

The self-adjoint Hamiltonian actually admits a broader class of boundary conditions at $r\to0$.
To see this, we express the Wronskian in Eq.~(\ref{eq:hermitian}) as
\begin{align}\label{eq:wronskian}
W[\varphi(r)^*,\chi(r)] = A[\varphi(r)]^*B[\chi(r)] - B[\varphi(r)]^*A[\chi(r)],
\end{align}
where
\begin{align}
A[\chi(r)] &\equiv W[\chi(r),f(r)\cos\delta - g(r)\sin\delta], \\[4pt]
B[\chi(r)] &\equiv W[\chi(r),f(r)\sin\delta + g(r)\cos\delta]
\end{align}
are defined with $f(r)$ and $g(r)$ being arbitrary real functions subject to $W[f(r),g(r)]=1$ and $\delta$ being an arbitrary real constant~\cite{Krall:1982}.
Therefore, if $\lim_{r\to0}A[\chi(r)]=0$ and $\lim_{r\to0}A[\varphi(r)]=0$ are imposed, the Hamiltonian is Hermitian and self-adjoint.
Although $\lim_{r\to0}B[\chi(r)]=0$ and $\lim_{r\to0}B[\varphi(r)]=0$ are also possible, they are redundant because of $B[\chi(r)]=A[\chi(r)]|_{\delta\to\delta-\pi/2}$.

In order for the boundary condition to allow nonvanishing solutions, it is appropriate to choose the auxiliary functions as two independent solutions to $\hat{H}f(r)=0$ and $\hat{H}g(r)=0$~\cite{Krall:1982}.
We then find
\begin{align}\label{eq:f(r)}
f(r) &= \sqrt{r}\,I_1\!\left(2\sqrt{\frac{2r}{a_0}}\right), \\
g(r) &= -2\sqrt{r}\,K_1\!\left(2\sqrt{\frac{2r}{a_0}}\right)
\end{align}
for the repulsive Coulomb potential or
\begin{align}
f(r) &= \sqrt{r}\,J_1\!\left(2\sqrt{\frac{2r}{a_0}}\right), \\
g(r) &= \pi\sqrt{r}\,Y_1\!\left(2\sqrt{\frac{2r}{a_0}}\right)
\label{eq:g(r)}\end{align}
for the attractive Coulomb potential in terms of the regular and singular modified Bessel or Bessel functions~\cite{DLMF}.
Because $\lim_{r\to0}A[\chi(r)]=0$ implies $\chi(r)$ proportional to their specific superposition of $f(r)\cot\delta-g(r)$ at $r\to0$, the boundary condition reads
\begin{subequations}\label{eq:boundary}
\begin{align}
\lim_{r\to0}\chi(r) \propto 1 + \left[\cot\delta
\pm \ln\!\left(e^{2\gamma-1}\frac{2r}{a_0}\right)\right]\frac{2r}{a_0} + O(r^2\ln r),
\end{align}
which constitutes the generalization of the Bethe-Peierls boundary condition in the presence of a Coulomb potential.
With Euler's constant $\gamma=0.577215\dots$, the resulting boundary condition is parametrized by $\cot\delta\in(-\infty,\infty)$, which in comparison to Eq.~(\ref{eq:bethe-peierls}) may be denoted as
\begin{align}\label{eq:scattering-length}
\cot\delta = -\frac{a_0}{2\tilde{a}}.
\end{align}
\end{subequations}
Here, $\tilde{a}$ is the analog of the scattering length in the presence of a Coulomb potential, and Eq.~(\ref{eq:bethe-peierls}) is indeed recovered from Eq.~(\ref{eq:boundary}) in the limit of $a_0\to\infty$ where the Coulomb potential disappears and $\tilde{a}\to a$.

\subsection{Relation to the effective-range expansion}
It is instructive to clarify our zero-range interaction in relation to the effective-range expansion in the presence of a Coulomb potential.
When a short-range potential exists only at $r\leq R$, the outer wave function solving the radial Schr\"odinger equation (\ref{eq:radial}) for $E=\hbar^2k^2/(2m)$ is generally provided by
\begin{align}\label{eq:solution}
\chi(r)|_{r>R} = C_\eta[F_\eta(\rho)\cot\tilde\delta(k) + G_\eta(\rho)].
\end{align}
Here, $F_\eta(\rho)\equiv F_0(\eta,\rho)$ and $G_\eta(\rho)\equiv G_0(\eta,\rho)$ are the regular and singular Coulomb wave functions in the $s$-wave channel, respectively, with $\rho=kr$, $\eta=\pm1/(ka_0)$, and $C_\eta=\sqrt{2\pi\eta/(e^{2\pi\eta}-1)}$~\cite{DLMF}, whereas $\tilde\delta(k)$ is the phase shift dependent on the short-range potential.
The effective-range expansion then reads
\begin{align}
C_\eta^2k\cot\tilde\delta(k) + 2k\eta h_\eta
= -\frac1{\tilde{a}} + \frac{\tilde{r}}{2}k^2 + O(k^4),
\end{align}
where $h_\eta=[\Psi(i\eta)+\Psi(-i\eta)]/2+\ln(ka_0)$ with $\Psi(z)$ being the digamma function and $\tilde{r}$ is the analog of the effective range in the presence of a Coulomb potential~\cite{Bethe:1949,Jackson:1950,Blatt-Weisskopf:1952}.

Since $R=0$ for the zero-range interaction, Eq.~(\ref{eq:solution}) extends down to $r\to0$, where the boundary condition in Eq.~(\ref{eq:boundary}) determines the phase shift as a function of $k$.
With the power-series expansions of the Coulomb wave functions in small $\rho$,%
\footnote{Specifically, $F_\eta(\rho)=C_\eta\rho+O(\rho^2)$ and $C_\eta G_\eta(\rho)=1+\eta\,[\Psi(i\eta)+\Psi(-i\eta)+2\ln(e^{2\gamma-1}2\rho)]\,\rho+O(\rho^2\ln\rho)$ for $\rho\to0$ are employed.}
we find
\begin{align}\label{eq:phase-shift}
C_\eta^2k\cot\tilde\delta(k) + 2k\eta h_\eta\Big|_\text{ZRI}
= -\frac1{\tilde{a}},
\end{align}
so that the effective range and all the higher-order coefficients vanish.
Therefore, the zero-range interaction (ZRI) is parametrized by the scattering length only as expected, which is also consistent with the causality bound requiring $\tilde{r}\leq0$ for $R=0$~\cite{Konig:2013}.

\subsection{Zero-range limit of a finite-range potential}
Although the boundary condition in Eq.~(\ref{eq:boundary}) is suitable for practical purposes, the zero-range interaction can also be implemented by taking the zero-range limit of a finite-range potential under fixed scattering length.
For demonstration, we employ a toy potential,
\begin{align}
V(r) =
\begin{cases}\displaystyle
-\frac{\hbar^2v^2}{2m} & (r\leq R), \\[8pt]\displaystyle
\pm\frac{\hbar^2}{ma_0r} & (r>R),
\end{cases}
\end{align}
where the Coulomb potential is replaced by a square-well potential at $r\leq R$.
The resulting scattering length can be determined by connecting the zero-energy solution $\chi(r)|_{r\leq R}\propto\sin(vr)$ to $\chi(r)|_{r>R}\propto f(r)\cot\delta-g(r)$ at $r=R$,%
\footnote{This is consistent with Eq.~(\ref{eq:solution}) and the effective-range expansion because of $\lim_{k\to0}C_\eta[F_\eta(\rho)\cot\tilde\delta(k)+G_\eta(\rho)]\propto f(r)\cot\delta-g(r)$.}
leading to
\begin{align}
v\cot(vR) = \frac{f'(R)\cot\delta-g'(R)}{f(R)\cot\delta-g(R)}
\end{align}
with $f(r)$ and $g(r)$ provided by Eqs.~(\ref{eq:f(r)})--(\ref{eq:g(r)}) and $\cot\delta$ by Eq.~(\ref{eq:scattering-length}).
The zero-range interaction is achieved by taking the limit of $R\to0$, where $\tilde{a}$ is fixed by tuning the potential depth according to
\begin{align}
v = \frac\pi{2R} + \frac2{\pi\tilde{a}}
\pm \frac{4}{\pi a_0}\ln\!\left(e^{-2\gamma}\frac{a_0}{2R}\right) + O(R\ln^2\!R)
\end{align}
at its least value.
Therefore, the zero-range interaction is always attractive, whose attraction becomes stronger (weaker) with increasing (decreasing) inverse scattering length, and $\tilde{a}$ depends not only on the short-range potential but also on the Coulomb potential.
In particular, $v=\pi/(2R)+2/(\pi a)+O(R)$ is recovered in the limit of $a_0\to\infty$, so that
\begin{align}
\frac1{\tilde{a}} = \frac1a
\mp \frac2{a_0}\ln\!\left(e^{-2\gamma}\frac{a_0}{2R}\right) + O(R\ln^2\!R)
\end{align}
holds for the square-well potential.

\section{Two charged particles}\label{sec:two-particles}
\subsection{Scattering matrix}
Let us apply the zero-range theory developed in the previous section to reveal bound states and resonances of two charged particles resulting as poles in the complex energy plane of the $S$ matrix.
To this end, we start with the wave function solving the radial Schr\"odinger equation (\ref{eq:radial}) for $E=\hbar^2k^2/(2m)$,
\begin{align}
\chi(r) = C_\eta[F_\eta(\rho)\cot\tilde\delta(k) + G_\eta(\rho)],
\end{align}
with the boundary condition in Eq.~(\ref{eq:boundary}) leading to the phase shift already determined as Eq.~(\ref{eq:phase-shift}).
Its asymptotic form at large $\rho$ is found to be
\begin{align}
\lim_{r\to\infty}\chi(r) &\propto e^{-i[\rho-\eta\ln(2\rho)]} \notag\\
&\quad - e^{2i[\sigma_\eta+\tilde\delta(k)]}e^{i[\rho-\eta\ln(2\rho)]} + O(\rho^{-1}),
\end{align}
where $\sigma_\eta=[\ln\Gamma(1+i\eta)-\ln\Gamma(1-i\eta)]/(2i)$ is the Coulomb phase shift.%
\footnote{Here, $F_\eta(\rho)=\sin[\rho-\eta\ln(2\rho)+\sigma_\eta]+O(\rho^{-1})$ and $G_\eta(\rho)=\cos[\rho-\eta\ln(2\rho)+\sigma_\eta]+O(\rho^{-1})$ for $\rho\to\infty$ are employed~\cite{DLMF}.}
The amplitude of the outgoing wave with respect to the incoming wave defines the $S$ matrix $S(k)=e^{2i[\sigma_\eta+\tilde\delta(k)]}$~\cite{Newton:1982}, which can be expressed as
\begin{align}
S(k) = \frac{\Gamma(1+i\eta)}{\Gamma(1-i\eta)}
\frac{-\frac1{\tilde{a}} - 2k\eta h_\eta + C_\eta^2ik}
{-\frac1{\tilde{a}} - 2k\eta h_\eta - C_\eta^2ik}
\end{align}
in terms of the Bohr radius and the scattering length.
Furthermore, the identities\,%
\footnote{They can be obtained with $\Psi(z)=\Psi(1+z)-1/z$ and $\Psi(z)=\Psi(1-z)-\pi\cot(\pi z)$~\cite{DLMF}.}
\begin{align}
h_\eta &= \Psi(1-i\eta) + \ln(+ika_0) + \frac{1-C_\eta^2}{2i\eta} \\
&= \Psi(1+i\eta) + \ln(-ika_0) - \frac{1-C_\eta^2}{2i\eta}
\label{eq:identity}\end{align}
lead to
\begin{align}\label{eq:s-matrix}
S(k) = \frac{\Gamma(1+i\eta)}{\Gamma(1-i\eta)}
\frac{-\frac1{\tilde{a}} - 2k\eta[\Psi(1-i\eta) + \ln(+ika_0)] + ik}
{-\frac1{\tilde{a}} - 2k\eta[\Psi(1+i\eta) + \ln(-ika_0)] - ik}
\end{align}
with $\eta=\pm1/(ka_0)$ for repulsive (upper sign) and attractive (lower sign) Coulomb potentials.

Whereas $|S(k)|=1$ holds for real $k$, the $S$ matrix analytically continued to complex $k$ may have poles~\cite{Taylor:1972,Newton:1982}.
In particular, poles at $\Re k=0$ with $\Im k>0$ ($\Im k<0$) correspond to bound (virtual) states whose wave functions decay (grow) exponentially at $r\to\infty$.
On the other hand, poles at $\Im k<0$ with $\Re k>0$ ($\Re k<0$) are called resonances (antiresonances) with complex energies, which always appear in pairs because of $S(k)^*=S(-k^*)$.

We note that $\Gamma(1+i\eta)$ has poles at $i\eta=-n$ with $n=1,2,3,\dots$ corresponding to infinite virtual (bound) states for a repulsive (attractive) Coulomb potential with no short-range potential.
They are not, however, poles of the $S$ matrix in Eq.~(\ref{eq:s-matrix}) because $\lim_{i\eta\to-n}\Gamma(1+i\eta)/\Psi(1+i\eta)=(-1)^n/\Gamma(n)$ is finite.
Similarly, $\Psi(1-i\eta)$ does not produce poles of the $S$ matrix.
Therefore, the $S$ matrix has poles only at $k$ satisfying
\begin{align}\label{eq:poles}
ik + \frac1{\tilde{a}} + 2k\eta[\Psi(1+i\eta) + \ln(-ika_0)] = 0,
\end{align}
with $k=i/a$ recovered in the limit of $a_0\to\infty$ where the Coulomb potential disappears and $\tilde{a}\to a$.
Its solutions in the complex $k$ plane are now analyzed as functions of $a_0/\tilde{a}\in(-\infty,\infty)$ for repulsive and attractive Coulomb potentials separately.
In particular, $\arg k\in(-\pi/2,3\pi/2)$ is considered because virtual states are not allowed by Eq.~(\ref{eq:poles}).

\begin{figure}[t]
\includegraphics[width=0.9\columnwidth]{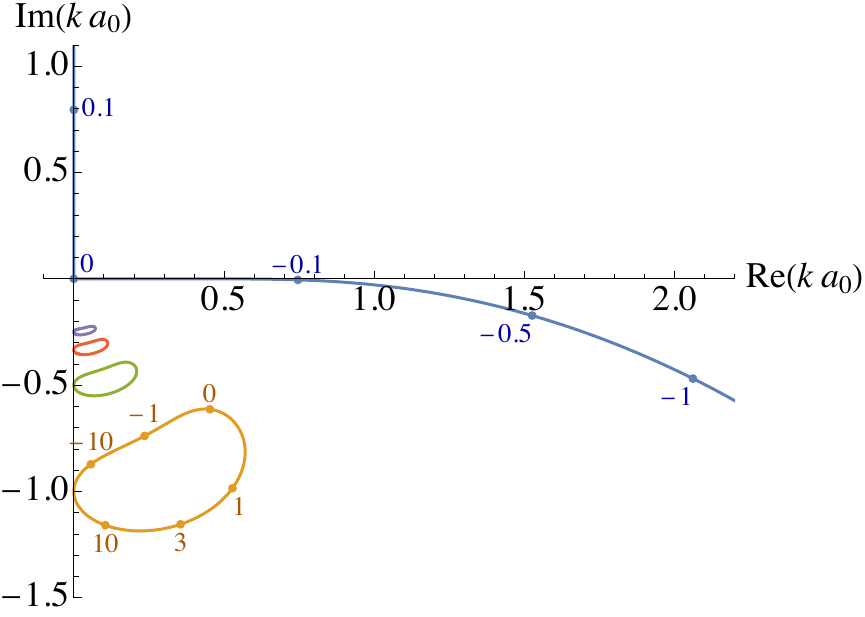}
\caption{\label{fig:repulsive}
Complex energies $E=\hbar^2k^2/(2m)$ of the lowest four and extra resonances in the complex plane of $ka_0$ as functions of the inverse scattering length $a_0/\tilde{a}$ normalized by the Bohr radius.
Several values of $a_0/\tilde{a}$ are indicated on their trajectories, whereas associated antiresonances on the side of $\Re k<0$ are not shown.}
\end{figure}

\subsection{Repulsive Coulomb potential}\label{sec:repulsive}
For a repulsive Coulomb potential, Eq.~(\ref{eq:poles}) with $\eta=+1/(ka_0)$ reads
\begin{align}\label{eq:repulsive}
ik + \frac1{\tilde{a}}
+ \frac2{a_0}\left[\Psi\!\left(1+\frac{i}{ka_0}\right) + \ln(-ika_0)\right] = 0,
\end{align}
which is found to support infinite resonances and antiresonances at arbitrary scattering length.
Their complex energies in the form of complex $k$ are shown in Fig.~\ref{fig:repulsive} as functions of $a_0/\tilde{a}$, where two distinct classes of solutions can be seen.
One class consists of infinite resonances with $n=1,2,3,\dots$, which appear out of the virtual Rydberg levels and eventually converge into the same levels as
\begin{align}
k = -\frac{i}{na_0} - \frac{2i\tilde{a}}{n^2a_0(a_0-2\pi i\tilde{a})} + O(\tilde{a}^2)
\end{align}
in the limits of $a_0/\tilde{a}\to\pm\infty$.
The other class consists of only one resonance, which appears out of
\begin{align}
k = \frac{i}{\tilde{a}} + \frac{2i}{a_0}\ln\left(-e^{-\gamma}\frac{a_0}{\tilde{a}}\right)
+ \frac{2\pi}{a_0} + O(\tilde{a}\ln\tilde{a})
\end{align}
in the limit of $a_0/\tilde{a}\to-\infty$ and then turns into a bound state right at $a_0/\tilde{a}\to0$,%
\footnote{Its wave function is normalizable even at $a_0/\tilde{a}=0$ and is provided by $\chi(r)=(4\sqrt{3r}/a_0)K_1(2\sqrt{2r/a_0})$.}
\begin{align}\label{eq:bound}
k = \sqrt{-\frac6{a_0\tilde{a}}} + O(\tilde{a}^{-3/2}),
\end{align}
whose binding energy eventually diverges as
\begin{align}
k = \frac{i}{\tilde{a}} + \frac{2i}{a_0}\ln\left(e^{-\gamma}\frac{a_0}{\tilde{a}}\right)
+ O(\tilde{a}\ln\tilde{a})
\end{align}
in the limit of $a_0/\tilde{a}\to\infty$~\cite{Schmickler:2019}.

\begin{figure}[t]
\includegraphics[width=0.9\columnwidth]{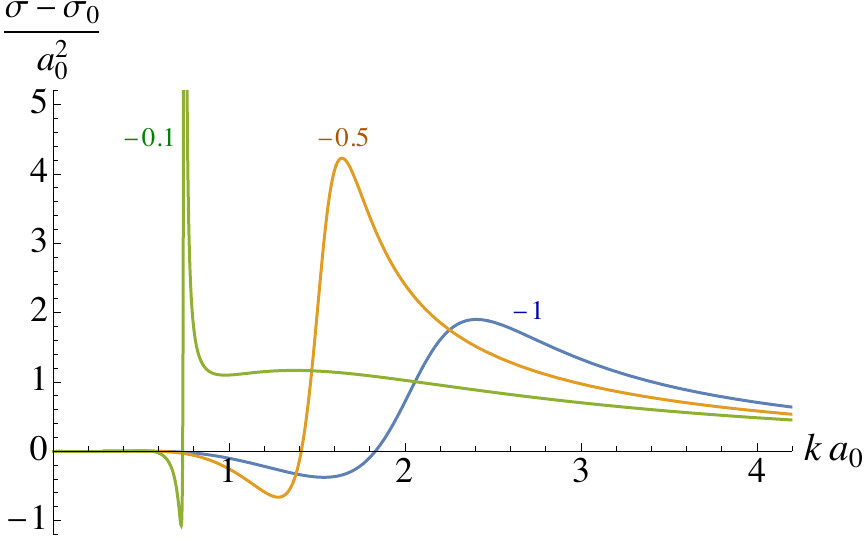}
\caption{\label{fig:scattering}
Scattering cross sections in Eq.~(\ref{eq:cross-section}) with $\sigma_0(k)=\pi|e^{2i\sigma_\eta}-1|^2/k^2$ subtracted as functions of $k=\sqrt{2mE/\hbar^2}$ normalized by the Bohr radius $a_0$.
The three curves correspond to different inverse scattering lengths $a_0/\tilde{a}=-1$, $-0.5$, and $-0.1$ as indicated by the plot labels.}
\end{figure}

Such a narrow resonance with $-\Im k\ll\Re k$ realized for $-1\lesssim a_0/\tilde{a}<0$ may be observable as a characteristic structure in the scattering cross section~\cite{Taylor:1972,Newton:1982}, which in the $s$-wave channel is provided by
\begin{align}\label{eq:cross-section}
\sigma(k) = \frac\pi{k^2}|S(k)-1|^2.
\end{align}
The resulting cross sections with respect to that with no short-range potential are shown in Fig.~\ref{fig:scattering} as functions of $ka_0$ for several choices of $a_0/\tilde{a}=-1$, $-0.5$, and $-0.1$.
With increasing inverse scattering length on its negative side, a sharp rise of the cross section is indeed developed at the resonance energy.

The repulsive Coulomb plus attractive short-range potentials are relevant to nuclear physics, where protons and nuclei are positively charged and interact via nuclear potentials.
Scattering lengths extracted from experimental data for various two-body systems are presented in Table~2 of Ref.~\cite{Konig:2013}.
For example, two protons have $a_0\approx57.6$~fm and $\tilde{a}\approx-7.83\pm0.01$~fm~\cite{Naisse:1977}, which are larger than the typical range of nuclear potential $R\approx1.4$~fm set by the inverse pion mass, but their large ratio $a_0/\tilde{a}\approx-7.36$ predicts no observable resonance.
On the other hand, two $\alpha$ particles have $a_0\approx3.63$~fm and extraordinary $\tilde{a}\approx(-1.92\pm0.09)\times10^3$~fm~\cite{Higa:2008}, which implies an observable resonance just above the threshold.
If the zero-range theory is naively applied, Eq.~(\ref{eq:bound}) with $a_0/\tilde{a}\approx-0.0019$ predicts the resonance at $E\approx9.0$~keV, which deviates from the measured energy of $E\approx184.1\pm0.1$~keV for the $^8$Be ground state by a factor of 20~\cite{Wustenbecker:1992}.
This is because the Bohr radius does not satisfy $a_0\gg R$, so that the effective range and possibly the higher-order coefficients are nonnegligible.
In particular, with the effective range of $\tilde{r}\approx1.1$~fm included into Eq.~(\ref{eq:repulsive}) by the replacement of $1/\tilde{a}\to1/\tilde{a}-\tilde{r}k^2/2$~\cite{Higa:2008}, the resonance energy is shifted up to $E\approx88$~keV, reducing the deviation down to a factor of 2.

Furthermore, a recent lattice QCD simulation shows that triply charmed baryons have $a_0\approx2.8$~fm, $\tilde{a}\approx-19$~fm, and $\tilde{r}\approx0.45$~fm~\cite{Lyu:2021}, which again implies an observable resonance just above the threshold.
If the zero-range theory is naively applied, Eq.~(\ref{eq:repulsive}) with $a_0/\tilde{a}\approx-0.15$ predicts the resonance at $E\approx(0.81-0.03\,i)$~MeV or $E\approx(1.3-0.2\,i)$~MeV without or with the effective range included.

\begin{figure}[t]
\includegraphics[width=0.9\columnwidth]{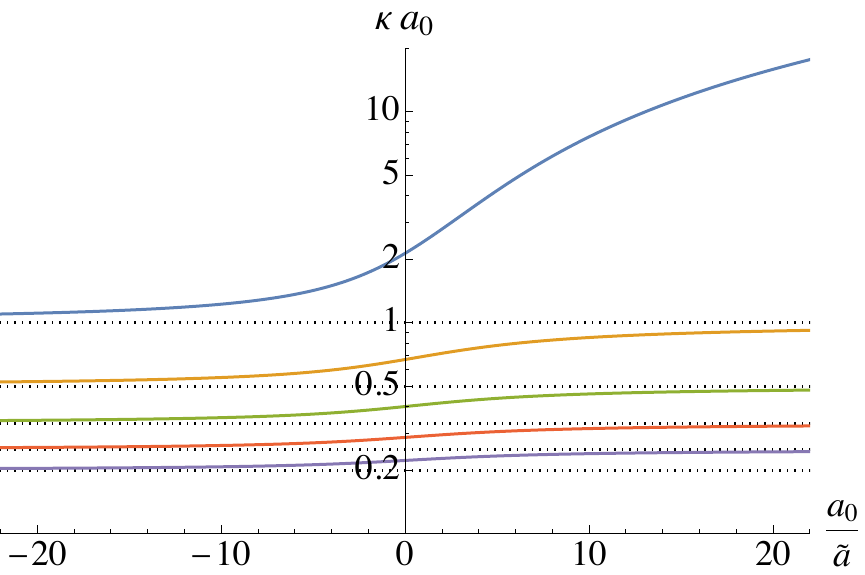}
\caption{\label{fig:attractive}
Binding energies $E=-\hbar^2\kappa^2/(2m)$ of the lowest five bound states in the form of $\kappa a_0$ as functions of the inverse scattering length $a_0/\tilde{a}$ normalized by the Bohr radius.
The horizontal dotted lines at $\kappa a_0=1/n$ indicate the Rydberg levels for $n=1,2,\dots,5$.}
\end{figure}

\subsection{Attractive Coulomb potential}
For an attractive Coulomb potential, Eq.~(\ref{eq:poles}) with $\eta=-1/(ka_0)$ reads
\begin{align}\label{eq:attractive}
ik + \frac1{\tilde{a}}
- \frac2{a_0}\left[\Psi\!\left(1-\frac{i}{ka_0}\right) + \ln(-ika_0)\right] = 0,
\end{align}
which is found to support infinite bound states at arbitrary scattering length.
Their binding energies are increasing functions of the inverse scattering length, which in the form of $\kappa=-ik>0$ are shown in Fig.~\ref{fig:attractive} as functions of $a_0/\tilde{a}$.
Here, all the solutions appear out of the Rydberg levels with $n=1,2,3,\dots$,
\begin{align}
\kappa = \frac1{na_0} - \frac{2\tilde{a}}{n^2a_0^2} + O(\tilde{a}^2)
\end{align}
in the limit of $a_0/\tilde{a}\to-\infty$.
Eventually, the ground-state energy ($n=1$) diverges as
\begin{align}
\kappa = \frac1{\tilde{a}} - \frac2{a_0}\ln\left(e^{-\gamma}\frac{a_0}{\tilde{a}}\right)
+ O(\tilde{a}\ln\tilde{a})
\end{align}
in the limit of $a_0/\tilde{a}\to\infty$, whereas each excited-state energy ($n=2,3,\dots$) converges into one lower Rydberg level as
\begin{align}
\kappa = \frac1{(n-1)a_0} - \frac{2\tilde{a}}{(n-1)^2a_0^2} + O(\tilde{a}^2).
\end{align}
We also note that the binding energies of higher excited states for $n\gg1$ are well approximated by
\begin{align}
\kappa \simeq \frac1{\left[n
- \frac1\pi\arccot\!\left(-\frac{a_0}{2\pi\tilde{a}}\right)\right]a_0}
\end{align}
with $\arccot(x)$ defined continuously in the range of $(0,\pi)$.
In contrast to such bound states, no resonances or antiresonances are found at any scattering length.

\section{Three charged particles}\label{sec:three-particles}
\subsection{Variational Born-Oppenheimer approximation}
The zero-range theory can be applied to more than two charged particles as well.
Because such systems tend to be intractable, we employ the Born-Oppenheimer approximation to study three equally charged particles.
When two of them are much heavier than the other, their positions can be fixed at $\R_1$ and $\R_2$, so that the light particle with its mass $m$ obeys $\hat{H}\psi(\r)=\eps\,\psi(\r)$ with
\begin{align}\label{eq:hamiltonian}
\hat{H} = -\frac{\hbar^2\nabla_{\!\r}^2}{2m} + \sum_{i=1,2}\frac{\hbar^2}{ma_0|\r-\R_i|}.
\end{align}
The resulting binding energy $\eps<0$ depends on the separation $R=|\R_1-\R_2|$ and is to serve as an effective potential between two heavy particles induced by the light particle.

In order to evaluate the induced effective potential, we assume a trial wave function,
\begin{align}\label{eq:trial}
\psi(\r) = C_\eta\left[\frac{H_\eta^+(\rho_1)}{\rho_1} + \frac{H_\eta^+(\rho_2)}{\rho_2}\right],
\end{align}
where $H_\eta^+(\rho)\equiv G_\eta(\rho)+iF_\eta(\rho)$ is the outgoing Coulomb wave function in the $s$-wave channel with $\rho_i=i\kappa|\r-\R_i|$ and $\eta=1/(i\kappa a_0)$~\cite{DLMF}.
Our wave function superposes two bound states localized at each heavy particle, which is motivated so as to become exact when two heavy particles are far separated or when the Coulomb potential disappears.
Its power-series expansion in small $\rho_1$ with the boundary condition in Eq.~(\ref{eq:boundary}) imposed at $\r\to\R_1$ leads to
\begin{align}
& -\kappa + \frac1{\tilde{a}}
+ \frac2{a_0}\left[\Psi\!\left(1+\frac1{\kappa a_0}\right) + \ln(\kappa a_0)\right] \notag\\
&\quad + \frac{C_\eta H_\eta^+(i\kappa R)}{R} = 0,
\end{align}
where the identity in Eq.~(\ref{eq:identity}) is also employed.
The last term is the correction due to the other zero-range interaction separated by $R$ and effectively adds to the attraction acting on the light particle.
Therefore, the resulting $\kappa$ increases compared to that from Eq.~(\ref{eq:repulsive}), which at $a_0/\tilde{a}=0$ is found to be
\begin{align}
\kappa|_{a_0/\tilde{a}=0} \simeq \frac{2\sqrt3\,\pi^{1/4}}{a_0}
\left(\frac{a_0}{2R}\right)^{3/8}\exp\biggl(-\sqrt{\frac{2R}{a_0}}\biggr)
\end{align}
in the limit of $R\to\infty$ and
\begin{align}
\kappa|_{a_0/\tilde{a}=0} \simeq \frac{c}{R}
\end{align}
in the limit of $R\to0$ with constant $c=0.567143\dots$ solving $c=e^{-c}$.%
\footnote{Here, $C_\eta H_\eta^+(r/\eta)\simeq 2\sqrt{2r}\,K_1(2\sqrt{2r})$ for $\kappa\to0$ and $C_\eta H_\eta^+(ir)\simeq e^{-r}$ for $\kappa\to\infty$ under fixed $r>0$ are employed.}

\begin{figure}[t]
\includegraphics[width=0.9\columnwidth]{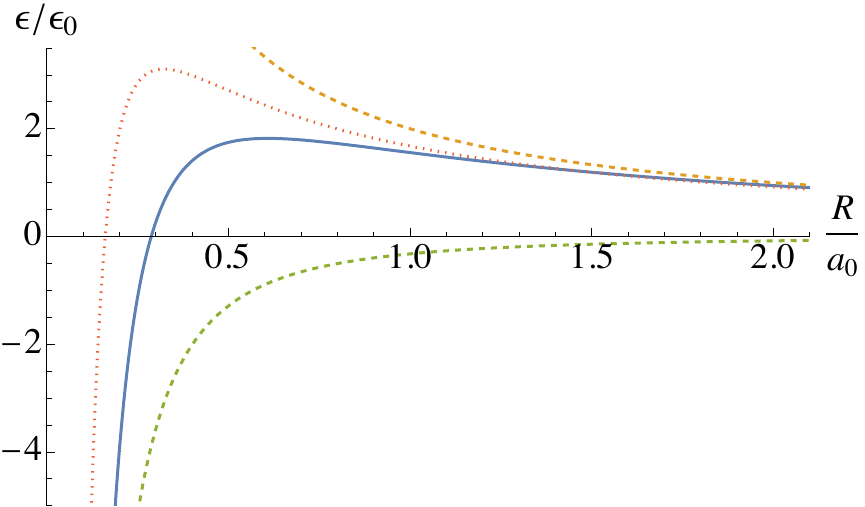}
\caption{\label{fig:potential}
Expectation value of the Hamiltonian in Eq.~(\ref{eq:hamiltonian}) with respect to the trial wave function in Eq.~(\ref{eq:trial}) as a function of $R/a_0$, where $\eps(R)=\<\psi|\hat{H}|\psi\>/\<\psi|\psi\>|_{a_0/\tilde{a}=0}$ is normalized by $\eps_0=\hbar^2/(2ma_0^2)$.
The upper and lower dashed curves plot $2a_0/R$ and $-(c\,a_0/R)^2$, respectively, which are $\eps(R)/\eps_0$ at $R\to\infty$ and $R\to0$.
Their sum $2a_0/R-(c\,a_0/R)^2$ is also plotted by the dotted curve.}
\end{figure}

The trial wave function in Eq.~(\ref{eq:trial}) approaches an eigenfunction of the Hamiltonian in Eq.~(\ref{eq:hamiltonian}) when two heavy particles are far separated.
Otherwise, its expectation value provides a variational upper bound on the ground-state energy, which at $a_0/\tilde{a}=0$ is shown in Fig.~\ref{fig:potential} as a function of $R/a_0$.
The resulting $\eps(R)=\<\psi|\hat{H}|\psi\>/\<\psi|\psi\>|_{a_0/\tilde{a}=0}$ has its asymptotic forms of
\begin{align}\label{eq:large-R}
\eps(R) \simeq \frac{\hbar^2}{ma_0R}
\end{align}
in the limit of $R\to\infty$ due to the repulsive Coulomb potential and
\begin{align}\label{eq:small-R}
\eps(R) \simeq -\frac{\hbar^2}{2m}\left(\frac{c}{R}\right)^2
\end{align}
in the limit of $R\to0$ due to the zero-range interaction.
Such an inverse-square attraction is known to lead to the Efimov effect in the absence of a Coulomb potential~\cite{Fonseca:1979}.

\subsection{Bound states and resonances}
The ground-state energy of the light particle may be regarded as an induced effective potential between two heavy particles.
In order to describe qualitative aspects of the system at $a_0/\tilde{a}=0$, we approximate $\eps(R)$ by a sum of its asymptotic forms at $R\to\infty$ and $R\to0$ in Eqs.~(\ref{eq:large-R}) and (\ref{eq:small-R}), respectively.
The relative wave function of two heavy particles with their mass $M\gg m$ then obeys
\begin{align}\label{eq:schrodinger}
\left[-\frac{\hbar^2}{M}\frac{d^2}{dR^2} - \frac{\hbar^2(1/4+s^2)}{MR^2}
+ \frac{2\hbar^2}{Ma_0'R}\right]\chi(R) = E\chi(R)
\end{align}
in the channel with angular momentum $L$.
Here, $1/4+s^2=Mc^2/(2m)-L(L+1)$ in the second term includes the centrifugal potential, whereas $Ma_0'=ma_0$ in the third term includes the Coulomb potential between two heavy particles, in addition to the contributions from the effective potential induced by the light particle.
In order to make the Hamiltonian self-adjoint, an appropriate boundary condition has to be imposed on the wave function at $R\to0$, which for the inverse-square attraction reads
\begin{align}\label{eq:inverse-square}
\lim_{R\to0}\chi(R) \propto \left[\Gamma(is)\left(\frac{\kappa_*R}{2}\right)^{1/2-is}
+ (s\to-s)\right] + O(R^{3/2}).
\end{align}
Here, the incoming and outgoing waves are superposed with the same amplitude whose relative phase is fixed by the three-body parameter $\kappa_*$ defined up to multiplicative factors of $e^{\pi/s}$~\cite{Braaten:2006}.

\begin{figure}[t]
\includegraphics[width=0.9\columnwidth]{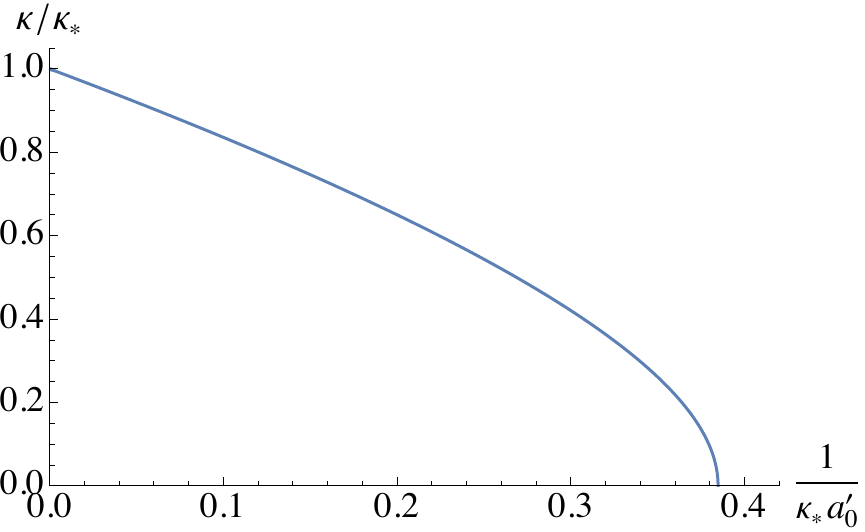}\\\smallskip
\includegraphics[width=0.96\columnwidth]{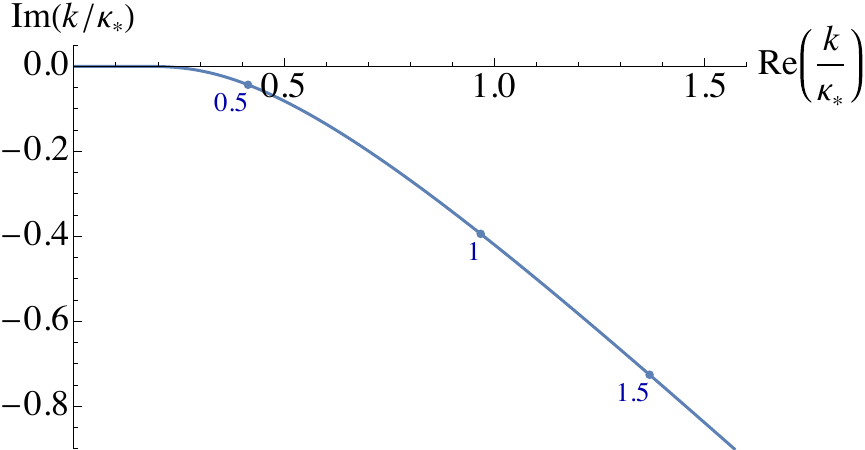}
\caption{\label{fig:bound-resonance}
Solution to Eq.~(\ref{eq:formula}) for $s=1$ in the form of $\kappa/\kappa_*$ or $k/\kappa_*$ with $k=i\kappa$ as a function of the inverse Bohr radius $1/(\kappa_*a_0')$ normalized by the three-body parameter.
The real solution in the upper panel turns into the complex solution in the lower panel at $1/(\kappa_*a_0')=0.3848\dots$.
Three values of $1/(\kappa_*a_0')$ are indicated on the trajectory of complex solution, whose associated solution $-k^*$ on the side of $\Re k<0$ is not shown.
Other infinite solutions are obtained by the replacement of $\kappa_*\to e^{-n\pi/s}\kappa_*$ with $n\in\Z$.}
\end{figure}

The normalizable solution to the radial Schr\"odinger equation (\ref{eq:schrodinger}) for $E=-\hbar^2\kappa^2/M<0$ is provided by
\begin{align}
\chi(R) = H_{-1/2+is}^+(\eta',i\kappa R),
\end{align}
where $H_l^+(\eta',\rho)$ is the outgoing Coulomb wave function with $\eta'=1/(i\kappa a_0')$ and angular momentum $l$~\cite{DLMF}.
Its power-series expansion in small $R$,
\begin{align}
\chi(R) &= (-i)^{-1/2+is+i\eta'}\sqrt{\frac{\Gamma\!\left(\frac12+is+i\eta'\right)}
{\Gamma\!\left(\frac12+is-i\eta'\right)}} \notag\\
&\quad \times \left[\frac{\Gamma(2is)\,(2\kappa R)^{1/2-is}}
{\Gamma\!\left(\frac12+is+i\eta'\right)} + (s\to-s)\right] + O(R^{3/2}),
\end{align}
with the boundary condition in Eq.~(\ref{eq:inverse-square}) imposed at $R\to0$ leads to\,%
\footnote{Our formula is also applicable to the case of an attractive Coulomb potential by the replacement of $a_0'\to-a_0'$, which is relevant to the problem studied in Ref.~\cite{Hammer:2008}.}
\begin{align}\label{eq:formula}
\left(\frac{\kappa}{\kappa_*}\right)^{2is}
= \frac{\Gamma\!\left(\frac12+is\right)\Gamma\bigl(\frac12-is+\frac1{\kappa a_0'}\bigr)}
{\Gamma\!\left(\frac12-is\right)\Gamma\bigl(\frac12+is+\frac1{\kappa a_0'}\bigr)}.
\end{align}
One of its solutions for real $\kappa$ as well as for complex $k=i\kappa$ under analytic continuation is shown in Fig.~\ref{fig:bound-resonance} as a function of $1/(\kappa_*a_0')$ by taking $s=1$ as an example.
The resulting $E=-\hbar^2\kappa^2/M$ or $E=\hbar^2k^2/M$ approximates the binding energy or complex resonance energy of three equally charged particles at infinite scattering length under our employed assumptions.

Infinite solutions are actually supported by Eq.~(\ref{eq:formula}) because if $\kappa/\kappa_*=(\kappa/\kappa_*)_\sol$ is a solution for $\kappa_*a_0'=(\kappa_*a_0')_\sol$, then $\kappa/\kappa_*=e^{-n\pi/s}(\kappa/\kappa_*)_\sol$ are also solutions for $\kappa_*a_0'=e^{n\pi/s}(\kappa_*a_0')_\sol$ with arbitrary $n\in\Z$.
In particular, infinite bound states with $\kappa=e^{-n\pi/s}\kappa_*$ obeying discrete scale invariance are recovered in the limit of $a_0'\to\infty$ where the Coulomb potential disappears.
Their binding energies are decreasing functions of the inverse Bohr radius and eventually vanish one by one at
\begin{align}
\left.\frac1{a_0'}\right|_{\kappa/\kappa_*=0}
= e^{-n\pi/s}\kappa_*\left[\frac{\Gamma\!\left(\frac12+is\right)}
{\Gamma\!\left(\frac12-is\right)}\right]^{1/(2is)},
\end{align}
where each bound state turns into a resonance by further increasing the inverse Bohr radius.

\section{Summary and outlook}\label{sec:summary}
In summary, we studied charged particles in three dimensions interacting via a short-range potential in addition to the Coulomb potential.
When the Bohr radius $a_0$ and the scattering length $\tilde{a}$ are much larger than the potential range $R$, low-energy (or long-wavelength $2\pi/k$) physics of the system becomes independent from details of the short-range potential.
The zero-range theory is suitable to describe such universal physics for $a_0,\tilde{a},k^{-1}\gg R$ directly in terms of the Bohr radius and the scattering length.
In particular, it was developed in this paper by generalizing the Bethe-Peierls boundary condition based on the self-adjoint extension of the Hamiltonian, which is applicable to arbitrary $N$ charged particles as well.
Namely, their wave function is made to obey the $N$-body Schr\"odinger equation without a short-range potential,
\begin{align}
\left[-\sum_{i=1}^N\frac{\hbar^2\nabla_i^2}{2m}
+ \sum_{i<j}\frac{\pm\hbar^2}{ma_0|\r_i-\r_j|}\right]\psi(\{\r\}) = E\psi(\{\r\}),
\end{align}
but with the boundary condition in Eq.~(\ref{eq:boundary}),
\begin{align}
\lim_{r=|\r_i-\r_j|\to0}\psi(\{\r\}) &\propto \frac1r - \frac1{\tilde{a}}
\pm \frac2{a_0}\ln\!\left(e^{2\gamma-1}\frac{2r}{a_0}\right) \notag\\
&\quad + O(r^2\ln r),
\end{align}
imposed whenever two coordinates come into contact.

The zero-range theory was then applied to two charged particles to reveal infinite resonances (bound states) for a repulsive (attractive) Coulomb potential and their trajectories in the complex energy plane.
It was also applied to three equally charged particles at infinite scattering length under the variational Born-Oppenheimer approximation, finding infinite bound states and resonances whose energies are fixed by the three-body parameter in addition to the Bohr radius.
Hopefully, our theoretical framework is to further reveal novel universal physics of few or many particles interacting via Coulomb plus short-range potentials.
Although realizing simultaneously large Bohr radius and scattering length may require accidental fine-tuning, resulting phenomena are potentially relevant to diverse systems in atomic, molecular, and chemical physics, nuclear and hadron physics, and dark-matter astrophysics due to the universality.

\acknowledgments
The authors thank Tetsuo Hyodo, Tomona Kinugawa, Daniel Phillips, and Takuma Yamashita for valuable discussions.
Our work was supported by JSPS KAKENHI Grant No.\ JP21K03384.


\begin{thebibliography}{99}

\bibitem{Braaten:2006}
E.~Braaten and H.-W.~Hammer,
``Universality in few-body systems with large scattering length,''
\href{https://doi.org/10.1016/j.physrep.2006.03.001}
{Phys.\ Rep.\ \textbf{428}, 259-390 (2006)}.

\bibitem{Bloch:2008}
I.~Bloch, J.~Dalibard, and W.~Zwerger,
``Many-body physics with ultracold gases,''
\href{https://doi.org/10.1103/RevModPhys.80.885}
{Rev.\ Mod.\ Phys.\ \textbf{80}, 885-964 (2008)}.

\bibitem{Giorgini:2008}
S.~Giorgini, L.~P.~Pitaevskii, and S.~Stringari,
``Theory of ultracold atomic Fermi gases,''
\href{https://doi.org/10.1103/RevModPhys.80.1215}
{Rev.\ Mod.\ Phys.\ \textbf{80}, 1215-1274 (2008)}.

\bibitem{Zwerger:2012}
\textit{The BCS-BEC Crossover and the Unitary Fermi Gas,}
edited by W.~Zwerger,
\href{https://doi.org/10.1007/978-3-642-21978-8}
{Lecture Notes in Physics, Vol.\ 836 (Springer, Berlin, 2012)}.

\bibitem{Randeria:2014}
M.~Randeria and E.~Taylor,
``Crossover from Bardeen-Cooper-Schrieffer to Bose-Einstein condensation and the unitary Fermi gas,''
\href{https://doi.org/10.1146/annurev-conmatphys-031113-133829}
{Annu.\ Rev.\ Condens.\ Matter Phys.\ \textbf{5}, 209-232 (2014)}.

\bibitem{Gandolfi:2015}
S.~Gandolfi, A.~Gezerlis, and J.~Carlson,
``Neutron matter from low to high density,''
\href{https://doi.org/10.1146/annurev-nucl-102014-021957}
{Annu.\ Rev.\ Nucl.\ Part.\ Sci.\ \textbf{65}, 303-328 (2015)}.

\bibitem{Horikoshi:2019}
M.~Horikoshi and M.~Kuwata-Gonokami,
``Cold atom quantum simulator for dilute neutron matter,''
\href{https://doi.org/10.1142/S0218301319300017}
{Int.\ J.\ Mod.\ Phys.\ E \textbf{28}, 1930001 (2019)}.

\bibitem{Nielsen:2001}
E.~Nielsen, D.~V.~Fedorov, A.~S.~Jensen, and E.~Garrido,
``The three-body problem with short-range interactions,''
\href{https://doi.org/10.1016/S0370-1573(00)00107-1}
{Phys.\ Rep.\ \textbf{347}, 373-459 (2001)}.

\bibitem{Jensen:2004}
A.~S.~Jensen, K.~Riisager, D.~V.~Fedorov, and E.~Garrido,
``Structure and reactions of quantum halos,''
\href{https://doi.org/10.1103/RevModPhys.76.215}
{Rev.\ Mod.\ Phys.\ \textbf{76}, 215-261 (2004)}.

\bibitem{Braaten:2007}
E.~Braaten and H.-W.~Hammer,
``Efimov physics in cold atoms,''
\href{https://doi.org/10.1016/j.aop.2006.10.011}
{Ann.\ Phys.\ \textbf{322}, 120-163 (2007)}.

\bibitem{Hammer:2010}
H.-W.~Hammer and L.~Platter,
``Efimov states in nuclear and particle physics,''
\href{https://doi.org/10.1146/annurev.nucl.012809.104439}
{Annu.\ Rev.\ Nucl.\ Part.\ Sci.\ \textbf{60}, 207-236 (2010)}.

\bibitem{Ferlaino:2011}
F.~Ferlaino, A.~Zenesini, M.~Berninger, B.~Huang, H.-C.~N\"agerl, and R.~Grimm,
``Efimov resonances in ultracold quantum gases,''
\href{https://doi.org/10.1007/s00601-011-0260-7}
{Few-Body Syst.\ \textbf{51}, 113-133 (2011)}.

\bibitem{Naidon:2017}
P.~Naidon and S.~Endo,
``Efimov physics: a review,''
\href{https://doi.org/10.1088/1361-6633/aa50e8}
{Rep.\ Prog.\ Phys.\ \textbf{80}, 056001 (2017)}.

\bibitem{Greene:2017}
C.~H.~Greene, P.~Giannakeas, and J.~P\'erez-R\'ios,
``Universal few-body physics and cluster formation,''
\href{https://doi.org/10.1103/RevModPhys.89.035006}
{Rev.\ Mod.\ Phys.\ \textbf{89}, 035006 (2017)}.

\bibitem{D'Incao:2018}
J.~P.~D'Incao,
``Few-body physics in resonantly interacting ultracold quantum gases,''
\href{https://doi.org/10.1088/1361-6455/aaa116}
{J.\ Phys.\ B: At.\ Mol.\ Opt.\ Phys.\ \textbf{51}, 043001 (2018)}.

\bibitem{Efimov:1970}
V.~Efimov,
``Energy levels arising from resonant two-body forces in a three-body system,''
\href{https://doi.org/10.1016/0370-2693(70)90349-7}
{Phys.\ Lett.\ B \textbf{33}, 563-564 (1970)}.

\bibitem{Efimov:1973}
V.~Efimov,
``Energy levels of three resonantly interacting particles,''
\href{https://doi.org/10.1016/0375-9474(73)90510-1}
{Nucl.\ Phys.\ A \textbf{210}, 157-188 (1973)}.

\bibitem{Kraemer:2006}
T.~Kraemer, M.~Mark, P.~Waldburger, J.~G.~Danzl, C.~Chin, B.~Engeser, A.~D.~Lange, K.~Pilch, A.~Jaakkola, H.-C.~N\"agerl, and R.~Grimm,
``Evidence for Efimov quantum states in an ultracold gas of caesium atoms,''
\href{https://doi.org/10.1038/nature04626}
{Nature (London) \textbf{440}, 315-318 (2006)}.

\bibitem{Nishida:2013}
Y.~Nishida, Y.~Kato, and C.~D.~Batista,
``Efimov effect in quantum magnets,''
\href{https://doi.org/10.1038/nphys2523}
{Nat.\ Phys.\ \textbf{9}, 93-97 (2013)}.

\bibitem{Kato:2020}
Y.~Kato, S.-S.~Zhang, Y.~Nishida, and C.~D.~Batista,
``Magnetic field induced tunability of spin Hamiltonians: Resonances and Efimov states in Yb$_2$Ti$_2$O$_7$,''
\href{https://doi.org/10.1103/PhysRevResearch.2.033024}
{Phys.\ Rev.\ Res.\ \textbf{2}, 033024 (2020)}.

\bibitem{Nakayama:2021}
Y.~Nakayama and Y.~Nishida,
``Efimov effect at the Kardar-Parisi-Zhang roughening transition,''
\href{https://doi.org/10.1103/PhysRevE.103.012117}
{Phys.\ Rev.\ E \textbf{103}, 012117 (2021)}.

\bibitem{Huang:1957}
K.~Huang and C.~N.~Yang,
``Quantum-mechanical many-body problem with hard-sphere interaction,''
\href{https://doi.org/10.1103/PhysRev.105.767}
{Phys.\ Rev.\ \textbf{105}, 767-775 (1957)}.

\bibitem{Olshanii:2001}
M.~Olshanii and L.~Pricoupenko,
``Rigorous approach to the problem of ultraviolet divergencies in dilute Bose gases,''
\href{https://doi.org/10.1103/PhysRevLett.88.010402}
{Phys.\ Rev.\ Lett.\ \textbf{88}, 010402 (2001)}.

\bibitem{Bethe:1935}
H.~Bethe and R.~Peierls,
``Quantum theory of the diplon,''
\href{https://doi.org/10.1098/rspa.1935.0010}
{Proc.\ R.\ Soc.\ London, Ser.\ A \textbf{148}, 146-156 (1935)}.

\bibitem{Domcke:1981}
W.~Domcke,
``Analytic theory of resonances, virtual states and bound states ion electron-molecule scattering and related processes,''
\href{https://doi.org/10.1088/0022-3700/14/24/022}
{J.\ Phys.\ B: At.\ Mol.\ Phys.\ \textbf{14}, 4889-4922 (1981)}.

\bibitem{Domcke:1983}
W.~Domcke,
``Analytic theory of resonances and bound states near Coulomb thresholds,''
\href{https://doi.org/10.1088/0022-3700/16/3/012}
{J.\ Phys.\ B: At.\ Mol.\ Phys.\ \textbf{16}, 359-380 (1983)}.

\bibitem{Florescu-Mitchell:2006}
A.~I.~Florescu-Mitchell and J.~B.~A.~Mitchell,
``Dissociative recombination,''
\href{https://doi.org/10.1016/j.physrep.2006.04.002}
{Phys.\ Rep.\ \textbf{430}, 277-374 (2006)}.

\bibitem{Lucchese:2019}
R.~R.~Lucchese, T.~N.~Rescigno, and C.~W.~McCurdy,
``The connection between resonances and bound states in the presence of a Coulomb potential,''
\href{https://doi.org/10.1021/acs.jpca.8b10715}
{J.\ Phys.\ Chem.\ A \textbf{123}, 82-95 (2019)}.

\bibitem{Bethe:1949}
H.~A.~Bethe,
``Theory of the effective range in nuclear scattering,''
\href{https://doi.org/10.1103/PhysRev.76.38}
{Phys.\ Rev.\ \textbf{76}, 38-50 (1949)}.

\bibitem{Jackson:1950}
J.~D.~Jackson and J.~M.~Blatt,
``The interpretation of low energy proton-proton scattering,''
\href{https://doi.org/10.1103/RevModPhys.22.77}
{Rev.\ Mod.\ Phys.\ \textbf{22}, 77-118 (1950)}.

\bibitem{Blatt-Weisskopf:1952}
J.~M.~Blatt and V.~F.~Weisskopf,
\textit{Theoretical Nuclear Physics}
(Wiley \& Sons, New York, 1952).

\bibitem{Taylor:1972}
J.~R.~Taylor,
\textit{Scattering Theory: The Quantum Theory on Nonrelativistic Collisions}
(Wiley \& Sons, New York, 1972).

\bibitem{Kong:1999}
X.~Kong and F.~Ravndal,
``Proton-proton scattering lengths from effective field theory,''
\href{https://doi.org/10.1016/S0370-2693(99)00144-6}
{Phys.\ Lett.\ B \textbf{450}, 320-324 (1999)};
\href{https://doi.org/10.1016/S0370-2693(99)00619-X}
{\textbf{458}, 565 (1999)}.

\bibitem{Kong:2000}
X.~Kong and F.~Ravndal,
``Coulomb effects in low energy proton–proton scattering,''
\href{https://doi.org/10.1016/S0375-9474(99)00406-6}
{Nucl.\ Phys.\ A \textbf{665}, 137-163 (2000)}.

\bibitem{Higa:2008}
R.~Higa, H.-W.~Hammer, and U.~van~Kolck,
``$\alpha\alpha$ scattering in halo effective field theory,''
\href{https://doi.org/10.1016/j.nuclphysa.2008.06.003}
{Nucl.\ Phys.\ A \textbf{809}, 171-188 (2008)}.

\bibitem{Konig:2013}
S.~K\"onig, D.~Lee, and H.-W.~Hammer,
``Causality constraints for charged particles,''
\href{https://doi.org/10.1088/0954-3899/40/4/045106}
{J.\ Phys.\ G: Nucl.\ Part.\ Phys.\ \textbf{40}, 045106 (2013)}.

\bibitem{Hammer:2017}
H.-W.~Hammer, C.~Ji, and D.~R.~Phillips,
``Effective field theory description of halo nuclei,''
\href{https://doi.org/10.1088/1361-6471/aa83db}
{J.\ Phys.\ G: Nucl.\ Part.\ Phys.\ \textbf{44}, 103002 (2017)}.

\bibitem{Hisano:2003}
J.~Hisano, S.~Matsumoto, and M.~M.~Nojiri,
``Unitarity and higher-order corrections in neutralino dark matter annihilation into two photons,''
\href{https://doi.org/10.1103/PhysRevD.67.075014}
{Phys.\ Rev.\ D \textbf{67}, 075014 (2003)}.

\bibitem{Hisano:2004}
J.~Hisano, S.~Matsumoto, and M.~M.~Nojiri,
``Explosive dark matter annihilation,''
\href{https://doi.org/10.1103/PhysRevLett.92.031303}
{Phys.\ Rev.\ Lett.\ \textbf{92}, 031303 (2004)}.

\bibitem{Hisano:2005}
J.~Hisano, S.~Matsumoto, M.~M.~Nojiri, and O.~Saito,
``Nonperturbative effect on dark matter annihilation and gamma ray signature from the galactic center,''
\href{https://doi.org/10.1103/PhysRevD.71.063528}
{Phys.\ Rev.\ D \textbf{71}, 063528 (2005)}.

\bibitem{Arkani-Hamed:2009}
N.~Arkani-Hamed, D.~P.~Finkbeiner, T.~R.~Slatyer, and N.~Weiner,
``A theory of dark matter,''
\href{https://doi.org/10.1103/PhysRevD.79.015014}
{Phys.\ Rev.\ D \textbf{79}, 015014 (2009)}.

\bibitem{Braaten:2017}
E.~Braaten, E.~Johnson, and H.~Zhang,
``Zero-range effective field theory for resonant wino dark matter.\ Part I.\ Framework,''
\href{https://doi.org/10.1007/JHEP11(2017)108}
{J.\ High Energy Phys.\ 11 (2017) 108}.

\bibitem{Braaten:2018a}
E.~Braaten, E.~Johnson, and H.~Zhang,
``Zero-range effective field theory for resonant wino dark matter.\ Part II.\ Coulomb resummation,''
\href{https://doi.org/10.1007/JHEP02(2018)150}
{J.\ High Energy Phys.\ 02 (2018) 150}.

\bibitem{Braaten:2018b}
E.~Braaten, E.~Johnson, and H.~Zhang,
``Zero-range effective field theory for resonant wino dark matter.\ Part III.\ Annihilation effects,''
\href{https://doi.org/10.1007/JHEP05(2018)062}
{J.\ High Energy Phys.\ 05 (2018) 62}.

\bibitem{Bonneau:2001}
G.~Bonneau, J.~Faraut, and G.~Valent,
``Self-adjoint extensions of operators and the teaching of quantum mechanics,''
\href{https://doi.org/10.1119/1.1328351}
{Am.\ J.\ Phys.\ \textbf{69}, 322-331 (2001)}.

\bibitem{Gitman-Tyutin-Voronov:2012}
D.~M.~Gitman, I.~V.~Tyutin, and B.~L.~Voronov,
\textit{Self-adjoint Extensions in Quantum Mechanics: General Theory and Applications to Schr\"odinger and Dirac Equations with Singular Potentials}
(Birkh\"auser, Boston, 2012).

\bibitem{Teschl:2014}
G.~Teschl,
\textit{Mathematical Methods in Quantum Mechanics: With Applications to Schr\"odinger Operators,}
2nd ed.\ (American Mathematical Society, Providence, 2014).

\bibitem{Krall:1982}
A.~M.~Krall,
``Boundary values for an eigenvalue problem with a singular potential,''
\href{https://doi.org/10.1016/0022-0396(82)90059-6}
{J.\ Diff.\ Equations \textbf{45}, 128-138 (1982)}.

\bibitem{DLMF}
NIST, ``Digital Library of Mathematical Functions,''
\url{https://dlmf.nist.gov/}.

\bibitem{Newton:1982}
R.~G.~Newton,
\textit{Scattering Theory of Waves and Particles,}
2nd ed.\ (Springer-Verlag, New York, 1982).

\bibitem{Schmickler:2019}
C.~H.~Schmickler, H.-W.~Hammer, and A.~G.~Volosniev,
``Universal physics of bound states of a few charged particles,''
\href{https://doi.org/10.1016/j.physletb.2019.135016}
{Phys.\ Lett.\ B \textbf{798}, 135016 (2019)}.

\bibitem{Naisse:1977}
J.~P.~Naisse,
``On precision analyses of the low-energy pp data,''
\href{https://doi.org/10.1016/0375-9474(77)90095-1}
{Nucl.\ Phys.\ A \textbf{278}, 506-524 (1977)}.

\bibitem{Wustenbecker:1992}
S.~W\"ustenbecker \textit{et al.,}
``Atomic effects on $\alpha$-$\alpha$ scattering to the $^8$Be ground state,''
\href{https://doi.org/10.1007/BF01291705}
{Z.\ Phys.\ A \textbf{344}, 205-217 (1992)}.

\bibitem{Lyu:2021}
Y.~Lyu, H.~Tong, T.~Sugiura, S.~Aoki, T.~Doi, T.~Hatsuda, J.~Meng, and T.~Miyamoto,
``Dibaryon with highest charm number near unitarity from lattice QCD,''
\href{https://doi.org/10.1103/PhysRevLett.127.072003}
{Phys.\ Rev.\ Lett.\ \textbf{127}, 072003 (2021)}.

\bibitem{Fonseca:1979}
A.~C.~Fonseca, E.~F.~Redish, and P.~E.~Shanley,
``Efimov effect in an analytically solvable model,''
\href{https://doi.org/10.1016/0375-9474(79)90189-1}
{Nucl.\ Phys.\ A \textbf{320}, 273-288 (1979)}.

\bibitem{Hammer:2008}
H.-W.~Hammer and R.~Higa,
``A model study of discrete scale invariance and long-range interactions,''
\href{https://doi.org/10.1140/epja/i2008-10617-3}
{Eur.\ Phys.\ J.\ A \textbf{37}, 193-200 (2008)}.

\end{thebibliography}
\end{document}